\documentclass[aps,preprint,nobibnotes,nofootinbib]{revtex4-1}
\usepackage{amssymb,amsfonts,amsmath}
\usepackage[all,arc]{xy}
\usepackage{enumerate}
\usepackage{mathrsfs}
\usepackage{graphicx}

\begin{document}

\title{Dependence of Self-force on Central Object}

\author{Theodore D. Drivas}
\author{Samuel E. Gralla}
\affiliation{\it Enrico Fermi Institute and Department of Physics \\ \it University of Chicago
\\ \it 5640 S.~Ellis Avenue, Chicago, IL~60637, USA}

\begin{abstract}
For a particle in orbit about a static spherically symmetric body, we study the change in self-force that results when the central body type (i.e., the choice of interior metric for the Schwarzschild exterior) is changed.  While a straight self-force is difficult to compute because of the need for regularization, such a ``self-force difference'' may be computed directly from the mode functions of the relevant wave equations.  This technique gives a simple probe of the (non)locality of the force, as well as offers the practical benefit of an easy determination of the self-force on a body orbiting an arbitrary (static spherically symmetric) central body, once the corresponding result for a black hole (or some other reference interior) is known.  We derive a general expression for the self-force difference at the level of a mode-sum in the case of a (possibly non-minimally coupled) scalar charge and indicate the generalization to the electromagnetic and gravitational cases.  We then consider specific choices of orbit and/or central body.  Our main findings are: (1) For charges held static at a large distance from the central body, the self-force is independent of the central body type in the minimally coupled scalar case and the electromagnetic case (but dependent in the nonminimally coupled scalar case); (2) For circular orbits about a thin-shell spacetime in the scalar case, the fractional change in self-force from a black hole spacetime is much larger for the radial (conservative) force than for the angular (dissipative) force; and (3) the radial self-force difference (between these spacetimes) agrees closely for a static charge and a circular orbit of the same radius.
\end{abstract}

\maketitle

\section{Introduction}

The leading order force exerted on a small body by its own self-field is known as the self-force.  Heuristically, this force can be broken up into two pieces.  One is local: the body's motion (and/or the properties of the medium it moves through) may distort the local self-field, causing an asymmetry over the body that exerts a net force.  The other is non-local: field ``sent out'' by the body may encounter the body again at a later time, resulting in a force that formally depends on the \textit{entire past history} of the particle motion.   This type of effect---called a ``tail'' effect---is unfamiliar because it does not occur for electromagnetic fields in empty four-dimensional flat spacetime, but it is fact generic: higher dimensional fields, massive fields, fields propagating on a curved spacetime, and fields propagating in the presence of non-trivial boundary conditions (such as caused by the presence of conductors) will generically give rise to non-local self-forces.

The above heuristic split may be made precise through the use of the Hadamard decomposition (e.g., \cite{poisson}) of a Green's function, which locally defines ``direct'' and ``tail'' pieces, supported on and interior to (respectively) the future light cone of the source point.  As demonstrated explicitly in the pioneering work of DeWitt and Brehme \cite{dewitt-brehme} in the context of a charged particle moving in a fixed curved background spacetime, the direct part gives rise to a local self-force, while the tail piece---together with contributions from the region outside the normal neighborhood, where the decomposition is not defined---gives rise to an integral of the derivative of the Green's function over the past history of the charge.  While we will consider one electromagnetic example in section \ref{sec:arb}, in this paper we will be primarily concerned with the case of a scalar charge first analyzed by Quinn \cite{quinn}, which shares many of the features of the electromagnetic case (and gravitational case) while avoiding much of the computational complexity.  Quinn considers the motion of a point particle coupled to a scalar field on a fixed globally hyperbolic background spacetime.  The spacetime is not required to be vacuum, but the use of the scalar wave equation (as opposed to some coupled matter and scalar field equation) entails the assumption that any matter present does not couple to the scalar field.  The spacetimes of this paper will contain matter only away from the particle, in which case Quinn's force becomes
\begin{equation}\label{sf}
F^\mu = q^2 \left[ \frac{1}{3} \left( \dot{a}^\mu - a^2 u^\mu \right) + \lim_{\epsilon \rightarrow 0^+} \int_{-\infty}^{\tau-\epsilon} \left( g^{\mu \nu} + u^\mu u^\nu \right) \nabla_\nu G(z(\tau),z(\tau')) d\tau' \right],
\end{equation}
where $G(x,x')$ is the retarded Green's function for the scalar wave equation, satisfying
\begin{equation}\label{greens}
\left[g^{\mu \nu}\nabla_\mu \nabla_\nu - \xi R\right]G(x,x')=-4\pi\delta^4(x,x')
\end{equation}
for some coupling constant $\xi$.  While Quinn did not include this coupling to curvature, his local analysis extends trivially to $\xi \neq 0$ when the particle moves through a vacuum region, as assumed here.  We include the curvature coupling in order to make equation \eqref{greens} more analogous to Maxwell's equation and the linearized Einstein equation, both of which involve couplings to curvature when expressed in wave-equation form (i.e., in the Lorenz gauge; c.f. \eqref{greens-EM} and e.g. \cite{poisson}).  In equation \eqref{sf} we take an explicitly perturbative viewpoint, where the particle's lowest-order motion $z^\mu(\tau)$ (four-velocity $u^\mu$, four-acceleration $a^\mu$, and four-jerk $\dot{a}^\mu$) is caused by an external scalar field (or other force), and the small self-force correction $F^\mu$ is viewed as a vector field defined on $z^\mu(\tau)$, to be used perturbatively or incorporated self-consistently in the spirit of \cite{gralla-wald, gralla-harte-wald}.  Note that since mass is not conserved for scalar charges, there is also a self-field effect on the mass \cite{quinn,harte}.  Since we view the scalar field primarily as a toy model for understanding the more physically interesting electromagnetic and gravitational cases (where mass is conserved), we do not compute this effect.  Similarly, we assume that the scalar charge $q$ is constant, although this is not required by the scalar theory \cite{quinn,harte,gralla}.

 The first term in equation \eqref{sf} is the local self-force coming from the direct part of the Green's function.  It takes an identical form to the classic Abraham-Lorentz-Dirac equation (c.f. \eqref{sf-EM} and e.g. \cite{poisson-ALD}) for electromagnetic fields, depending on the derivative of the acceleration.  The second term is the nonlocal self-force coming from the tail portion of the Green's function.\footnote{The limiting procedure ensures that the direct portion of the Green's function---which is nonintegrably singular---is not included.  See (e.g.) Poisson \cite{poisson} for details.}  Although this ``tail integral'' depends upon the entire past history of the particle, in physically reasonable situations one expects the contribution from the distant past to be negligible.  That is, there must be some point in the past at which the tail integral can be cut off, with numerically negligible error.  The question of just how far in the past one must integrate is the question of the \textit{locality} of the tail integral.

Several authors have already investigated this question \cite{anderson-wiseman,anderson-hu,anderson-flanagan-ottewill,cassals-dolan-ottewill-wardell}, primarily in the context of attempting to compute the tail integral via a series expansion of the Green's function in the normal neighborhood of the charge.  A generic finding of these authors is that the region of non-negligible support of the tail integral extends well beyond the normal neighborhood: the tail integral is too  non-local to be handled by series expansions alone.  This non-locality can also be seen in the case of a slow-moving particle in a spherically symmetric weakly-curved spacetime, in an approach originally due to deWitt and deWitt \cite{dewitt-dewitt} in the electromagnetic case and later improved and extended to the scalar and gravitational cases by Pfenning and Poisson \cite{pfenning-poisson}.  In these remarkable calculations, one considers a body that is freely falling to lowest order, so that the direct (local) portion of the self-force vanishes and only the tail (non-local) portion remains.  However, since by assumption this body moves slowly in a weakly curved spacetime, one should be able to regard the effects of curvature as a ``force'' causing a gravitational ``acceleration'' $\vec{g}$ defined relative to inertial motion in the flat background.  From this point of view one would expect the self-force to be given by the usual Abraham-Lorentz-Dirac type direct force, proportional to the time derivative of the ``acceleration'' $\vec{g}$.  Indeed, this result is recovered: when the tail integral is computed, it gives rise to exactly the Newtonian expectations.

However, this result comes with a twist: the tail integral that gives rise to the local Newtonian answer is in fact highly non-local!  Rather than it becoming dominated by a local contribution in the weak-field limit, as one might expect, it turns out that one must integrate at least a light reflection time (off the central body) in the past to recover the simple local self-force.  In fact, if one views the central body as a point mass, then in the scalar case---as clearly explained by Anderson and Wiseman \cite{anderson-wiseman}---the tail integrand actually \textit{vanishes} in the immediate past of the particle, giving its first contribution at a light reflection time off the point mass in the past.  This provides a dramatic illustration of the way in which a local expression for the force can arise from a highly non-local tail integral.  We will refer to such a self-force as \textit{weakly local}: it depends only on local properties, but the tail integral may be non-local.

A direct evaluation of the tail integral is not the only way to calculate a self-force.  Since the tail integral appears as part of an analytic expression for the field near a point particle, it is often (and even usually) more convenient to simply compute the field of a point particle by some other method and then extract the tail piece by a regularization procedure.  This type of approach gives no direct information on the locality of the integral, but by \textit{comparing} the results of different calculations one can glean some relevant information.  For example, consider two spacetimes that agree in a region of the charge but disagree elsewhere.  If there is a disagreement in the self-force on a particle moving through the region of agreement, then the tail must be at least as non-local as the light reflection time across the region.\footnote{More precisely, consider the causal diamond defined by the current particle position $z(\tau)$ and a prior particle position $z(\tau')$.  If this diamond lies within the region of agreement, then the Green's function $G(z(\tau),z(\tau'))$ must agree for both spacetimes.  Thus, any difference in the self-force must come from times $\tau'$ far enough in the past that the causal diamond intersects the central body.}  And if the results agree for a large class of locally-agreeing spacetimes, one would say that the self-force is weakly local in that case.

An example of this type of comparison was discussed by Pfenning and Poisson \cite{pfenning-poisson}, who compared their results for Newtonian central bodies with known results for black holes \cite{smith-will,wiseman}.  This comparison is conveniently summarized in table \ref{fig:sfresults}, which shows results for static charges in the far-field of Newtonian stars and black holes.
\begin{figure}[t]\label{fig:sfresults}
\begin{center}
  \begin{tabular}{ | l | c | r | }
    \hline
    \           & star SF & black hole SF \\ \hline
    scalar case & $F_r=2 \xi q^2 M/r^3$ & 0 \\ \hline
    EM case     & $F_r=e^2 M/r^3$ & $F_r=e^2 M/r^3$ \\
    \hline
  \end{tabular}
\end{center}
\caption{A table of known self-force results for static charges in the far field of Newtonian stars and black holes.  The Newtonian star results are taken from Pfenning and Poisson \cite{pfenning-poisson}, while the black hole results (which are exact if $r$ is the Schwarzschild coordinate) are due to Wiseman \cite{wiseman} in the scalar case and Smith and Will \cite{smith-will} the electromagnetic case.  Here $M$ is the mass of the central body, and $q$/$e$ is the scalar/electric charge of the particle.}
\end{figure}
We see that in the electromagnetic and minimally-coupled ($\xi=0$) scalar cases, the far-field self-force is independent of the central object type (Newtonian star or black hole), whereas for the nonminimally-coupled scalar case there is a difference.  This difference confirms the non-locality of the tail integral; but it is interesting that weak locality holds only in the electromagnetic and minimally coupled scalar cases.  What is so different about the nonminimally coupled scalar case?

In order to better understand as well as significantly extend these types of results, in this paper we make a direct attack on the dependence of the self-force on the central object.  In effect, we take the method of comparison to an extreme: rather than computing a self-force, we will directly compute the \textit{difference} in self-force that results from changing the interior metric while keeping the Schwarzschild exterior the same.  Such a ``self-force difference'' is easy to compute because the direct portions of the Green's functions cancel out upon subtraction, automatically picking out the relevant tail portion.  More precisely, for points $x$ and $x'$ within the region where the two spacetimes agree, define the difference $\Delta G(x,x')$ between the retarded Green's functions.  But initial value results imply that the Green's functions agree for points whose causal diamond lies within the region of agreement, so that in particular the difference $\Delta G$ must vanish in a neighborhood of coincidence, $x=x'$.  This means that when computing the difference $\Delta F^\mu$ in force on a particle moving through the region of agreement, we may actually drop the limit in equation \eqref{sf}, giving simply
\begin{equation}\label{sfdiff-greens}
\Delta F^\mu = q^2 \int \left( g^{\mu \nu} + u^\mu u^\nu \right) \nabla_\nu \Delta G(z(\tau),z(\tau')) d\tau'.
\end{equation}
Note that we have extended the worldline integral in the future direction, which is possible because of the vanishing of the support of the retarded Green's function for $x'$ in the future of $x$.  As we shall see, equation \eqref{sfdiff-greens} can be evaluated directly from the mode functions of the relevant spacetimes, avoiding the delicate regularization needed to compute a straight self-force, equation \eqref{sf}.

We consider static spherically symmetric bodies (described by a metric of the form \eqref{bodyMetric}) and obtain the following results.  First, we ``fill in the gaps'' between Newtonian stars and black holes in table \ref{fig:sfresults}, providing analogous results for arbitrary central bodies.  Specifically, we show that the far-field self-force on a static charge is indeed independent of the central body (i.e., weakly local) in the electromagnetic and minimally-coupled scalar cases, and that the force indeed depends in detail on the body type in the nonminimally coupled scalar case.  In this latter case we provide a specific example of the detailed dependence by considering a central body that consists of a thin-shell of radius $r_0$.  As one changes the radius $r_0$ between some very large radius (Newtonian object) and $r_0=2M$ (black hole), one can see the far-field force interpolate\footnote{However, the transition is not smooth everywhere, because instabilities can occur in some regions of parameter space---see appendix.} between the two cases (figure \ref{fig:staticplot}).  Our analysis also sheds some light on the reason the nonminimally coupled scalar case behaves differently from the other cases: weak locality in this limit is identified with the property that the static spherically symmetric solution to a wave equation is the same for all metrics.  This property holds for the minimally-coupled scalar and electromagnetic equations (the solution is simply the constant solution), but does not hold for the non-minimally coupled scalar wave equation.

We next move to the simplest dynamic case of circular orbits.  Here there is a dissipative self-force---the result of energy being radiated away from the system---in addition to the conservative effects seen in the static case.  We numerically compute the self-force difference between a thin-shell spacetime and Schwarzschild (figure \ref{fig:sfnumbers}), finding that the difference in interior is far more important for the conservative part of the force than for the dissipative part (figure \ref{fig:sffrac}).  This is in accord with the intuition that the dissipative self-force is more like a local radiation reaction force, whereas the conservative self-force should be sensitive to boundary conditions.\footnote{A simple example of a conservative self-force having nothing to do with curved spacetime is the force on a charged particle in the presence of a conductor.} This constitutes numerical evidence for weak locality of the dissipative force even in the strong-field regime, and holds out the possibility of a reasonable degree of true tail locality in this regime.\footnote{The idea that the tail may become \textit{more} local in the strong-field regime can be supported by the intuition that field is bent more strongly (and hence quickly) back to the particle in the presence of large curvatures.}  We also notice (figure \ref{fig:circstat}) that the conservative self-force difference closely matches the static case self-force difference (whereas the straight self-forces do not match).

Although this paper gives little quantitative handle on the precise degree of (non)locality of the tail, we take a moment to point out a potential practical application, should such knowledge ever be gained.  Much current effort is directed towards the production of waveform templates for extreme mass-ratio inspiral to be used in LISA \cite{lisa} data analysis, and here the gravitational self-force must take a central role (see \cite{barack} for a recent review).  The use of frequency-domain techniques---convenient for circular orbits---becomes much less feasible for very complicated orbits, where a great many frequencies would be needed to reconstruct the field.  However, suppose one knew that the tail integral was precisely $T$ local, in that one could cut off the integration at a time $T$ in the past with negligible error.  This means that in an indirect computation of the force, one can replace the particle orbit with \textit{anything one wants} in the region before time $T$ in the past, while not affecting the value of the self-force.  This freedom could potentially be exploited to greatly improve the convergence of the Fourier series used to reconstruct the field and force.  For example, a zoom-whirl orbit of Kerr could be replaced by a (non-geodesic) circular orbit of the whirl radius and frequency.  More generally, one could imagine having an algorithm to modify the past properties of orbits in order to maximize the estimated speed of convergence of the inverse fourier transform.

Finally, we note that the techniques used in this paper could be used to help generate waveform templates for the case where the central body is not a black hole.  For spinning objects, the exterior spacetime metric will depend on the central body, and any departure from a Kerr black hole will show up at lowest order in the motion of the inspiraling body.  But for non-spinning bodies (that are also roughly spherical), the exterior metric will remain Schwarzschild and differences in motion will be seen first in the self-force.  However, by the methods used in this paper, this self-force is easily computed for a given body type once the corresponding result for Schwarzschild is known.  Therefore, it appears that it should be a relatively computationally inexpensive matter to broaden search templates to include central objects that are not black holes, even in the non-spinning case.

In section \ref{sec:sssb} we derive a general frequency-domain expression for scalar self-force differences involving static, spherically symmetric central bodies.  In section \ref{sec:arb} we evaluate this expression (and its electromagnetic analog) for static charges in the far-field limit and extend the results of table \ref{fig:sfresults} to arbitrary central bodies.  Finally in section \ref{sec:shell} we consider a thin-shell central body and compute the self-force difference for static and circular orbits of arbitrary radius.  Our conventions are those of Wald \cite{wald-book}.

\section{General Expression for Spherically Symmetric Central Bodies}\label{sec:sssb}

To construct a metric representing a general static spherically symmetric body we will match a general interior metric form (see e.g. \cite{schutz}) with the Schwarzschild exterior.  Taking the matching surface to be at a radius $r_0>2M$, our spacetime will be
\begin{equation}
ds^2 = \begin{cases}\label{bodyMetric}
\ -e^{2\Psi(\tilde{r})} dt^2 + e^{2\Lambda(\tilde{r})} d\tilde{r}^2 + \tilde{r}^2 d\Omega^2 & \quad 0 < \tilde{r} < r_0 \\
\ -f(r) dt^2 + f(r)^{-1} dr^2 + r^2 d\Omega^2 & \quad r > r_0
\end{cases},
\end{equation}
where $f(r)=1-2M/r$ and $d\Omega^2 = d\theta^2 + \sin^2\theta d\phi^2$, and $\Psi$ and $\Lambda$ are analytic\footnote{The assumption of analyticity enables us to establish existence and uniqueness properties for the mode functions associated with the interior metric.} functions such that $e^{2\Psi(r_0)}=f(r_0)$.  The last requirement arises from the demand that the induced metric agree on the matching surface (so that the geometry is smooth enough to interpret as a single solution \cite{israel,poisson-book}), together with our choice to use the ``same'' coordinate $t$ on both sides of the matching surface.  However, this places no physical restrictions on the interior metric, since a rescaling of the time coordinate can always be used to rescale $\Psi$ so as to satisfy $e^{2\Psi(r_0)}=f(r_0)$.  We do not demand that the extrinsic curvature matches, which corresponds to allowing a surface layer of matter to be present at $r=r_0$ \cite{israel,poisson-book}.

While the metric components normal to the surface play no role in the matching analysis, it is often convenient to work with the same radial coordinate on both sides.  In this paper we will make the simple choice that
\begin{equation}\label{rrtilde}
\tilde{r} = r_0 + (r-r_0)e^{-\Lambda(r_0)}f(r_0)^{-1/2},
\end{equation}
given which the metric \eqref{bodyMetric} may be written in the single coordinate system $(t,r,\theta,\phi)$ as
\begin{equation}
ds^2 = \begin{cases}\label{bodyMetric2}
\ -e^{2\Psi(\tilde{r})} dt^2 + e^{2(\Lambda(\tilde{r})-\Lambda(r_0))}f(r_0)^{-1} dr^2 + \tilde{r}^2 d\Omega^2 & \quad r_0\left[1 - e^{\Lambda(r_0)}f(r_0)^{1/2}\right] < r < r_0 \\
\ -f(r) dt^2 + f(r)^{-1} dr^2 + r^2 d\Omega^2 & \quad r > r_0
\end{cases}.
\end{equation}
In these coordinates, all metric components agree at the matching surface $r=r_0$.

We will solve for the Green's function \eqref{greens} on the background \eqref{bodyMetric} by making use of separation of variables.  The time translation and spherical symmetries of the metric imply that we should decompose as follows,\footnote{There is no guarantee that this decomposition will always exist for the class of spacetimes \eqref{bodyMetric} we consider.  For example, the presence of instabilities could render the mode functions $G_{\ell\omega}$ ill-defined, as happens for the case discussed in the appendix.  We simply assume that this decomposition exists, in the hopes that this eliminates only unphysical spacetimes.}
\begin{equation}\label{greens-decomp}
G(x,x')=\int_{-\infty}^\infty d\omega \sum_{\ell=0}^{\infty}\sum_{m=-\ell}^ \ell Y_{\ell m}(\theta, \phi)  Y_{\ell m}^*(\theta', \phi')  G_{\ell\omega}(r,r')e^{-i\omega t}e^{i\omega t'},
\end{equation}
where $Y_{lm}$ are the spherical harmonics \cite{jackson}.
Note that we write $G_{l\omega}$ since the rotational symmetry also implies that the mode functions are independent of $m$ (as seen explicitly in equations \eqref{ArbWaveEqn} and \eqref{SchwWaveEqn} below).  After likewise decomposing the delta-function source into modes, equation \eqref{greens} implies
\begin{align}
\tilde{r}<r_0: \quad & G_{\ell\omega}''(\tilde{r}) +\left(-\Lambda'+\frac{2}{\tilde{r}}+\Psi'\right)G_{\ell\omega}'(\tilde{r}) - e^{2\Lambda} \left(\frac{\ell(\ell +1)}{\tilde{r}^2} - e^{-2\Psi} \omega^2 + \xi R \right) G_{\ell\omega}(\tilde{r})  = 0 \label{ArbWaveEqn}\\
r>r_0: \quad & G_{\ell\omega}''(r)+\frac{2(r-M)}{r^2 f}G_{\ell\omega}'(r)- \left(\frac{\ell(\ell +1)}{r^2f}- \frac{\omega^2}{f^2}\right)G_{\ell\omega}(r) = -\frac{4\pi}{r^2 f}\delta(r-r_q) \label{SchwWaveEqn},
\end{align}
where we have defined $r_q=r'$ to allow the prime to indicate ordinary differentiation.  (The notation $r_q$ reminds that the particle will be at the radius $r=r_q$, which we assume to be at $r>r_0$.) The solution for $G_{\ell \omega}$ is constructed by matching together homogeneous solutions from the regions $0<\tilde{r}<r_0$, $r_0<r<r_q$, and $r>r_q$.  To determine the matching conditions it is easiest to work with the same coordinate $r$ on both sides of $r_0$, i.e., to use the metric form \eqref{bodyMetric2}.  In this case the wave equation can be written in the form
\begin{align}
& G_{\ell\omega}''(r) + M(r) G_{\ell\omega}'(r) + \left[ N(r) + \xi f^{-1}\chi \delta(r-r_0) \right] G_{\ell\omega}(r) = \frac{-4\pi}{r^2 f}\delta(r-r_q),
\end{align}
where $N(r)$ is a continuous function, $M(r)$ is a continuous function except for a possible jump discontinuity at $r_0$, and $\chi$ is the coefficient of the delta-function contribution to the Ricci scalar (coming from a possible boundary layer of matter at $r_0$),
\begin{equation}
\chi = \lim_{\epsilon \rightarrow 0} \int_{r_0-\epsilon}^{r_0+\epsilon} R(r) dr.
\end{equation}
It is now easy to see that the matching conditions are simply
\begin{align}
\lim_{r\rightarrow r_0^+}G_{\ell\omega}(r)-\lim_{r\rightarrow r_0^-} G_{\ell\omega}(r) &= 0 \label{matchr0G} \\
\lim_{r\rightarrow r_0^+}G'_{\ell\omega}(r)-\lim_{r\rightarrow r_0^-} G'_{\ell\omega}(r) &= \left. f^{-1} \xi \chi G_{\ell\omega} \right|_{r=r_0} \label{matchr0Gp} \\
\lim_{r\rightarrow r_q^+}G_{\ell\omega}(r)-\lim_{r\rightarrow r_q^-} G_{\ell\omega}(r) &= 0 \label{matchrqG} \\
\lim_{r\rightarrow r_q^+}G'_{\ell\omega}(r)-\lim_{r\rightarrow r_q^-} G'_{\ell\omega}(r) &= \left. -\frac{4\pi}{r^2f}\right|_{r=r_q}. \label{matchrqGp}
\end{align}
Note that all derivatives are taken with respect to $r$ (rather than $\tilde{r}$) in the above equations.  We further impose the boundary conditions of regularity at $\tilde{r}=0$ and outgoing radiation ($G_{\ell \omega} \rightarrow r^{-1} e^{i \omega r}$) at $r \rightarrow \infty$, which should correspond to choosing the retarded Green's function, as demanded by the self-force prescription \eqref{sf}.
  
We will begin by performing the matching at $r=r_q$, which is well-studied since equation \eqref{SchwWaveEqn} is simply the Schwarzschild radial equation.  In treating this equation it is convenient to choose homogeneous solutions $R^H_{\ell\omega}$ and $R^\infty_{\ell\omega}$ defined by the end behavior\footnote{The forms given in equations \eqref{RH} and \eqref{RI} are not valid for $\omega=0$.  In this case, one simply demands finite limits, so that $R^H_{\ell 0}$ is regular at the horizon, and $R^\infty_{\ell 0}$ falls off at infinity.  In fact, $R^H_{\ell 0}$ and $R^\infty_{\ell 0}$ are simply Legendre functions---see section \ref{sec:shell}.}
\begin{align}
R^H_{\ell\omega}(r) &\rightarrow \frac{e^{-i\omega r_*}}{r} \ \ \ \ \ \ \ \text{for} \ \ \ \ r_* \rightarrow -\infty \label{RH} \\
R^\infty_{\ell\omega}(r) & \rightarrow \frac{e^{+i\omega r_*}}{r} \ \ \ \ \ \ \ \text{for}  \ \ \ \ r_* \rightarrow +\infty, \label{RI}
\end{align}
where $r*$ is the ``tortoise coordinate'' defined by $dr_*/dr = f^{-1}$ and $r* \rightarrow r$ as $r \rightarrow \infty$.  From equation \eqref{greens-decomp} we see that $R^H$ corresponds to purely ingoing radiation at the horizon,\footnote{Although the Schwarzschild metric does not extend to $r=2M$ in our spacetime, it is still convenient to choose homogeneous basis solutions to \eqref{SchwWaveEqn} with properties at $r=2M$ in mind.} while $R^\infty$ corresponds to purely outgoing radiation at infinity.  Matching these solutions at $r=r_q$ gives a useful particular solution to \eqref{SchwWaveEqn},
\begin{equation}\label{gschw}
G^{\textrm{Schw}}_{\ell\omega}=-N_{\ell\omega} R^H_{\ell\omega}(r_<) R^{\infty}_{\ell\omega}(r_>) , \qquad N_{\ell \omega} = \frac{4 \pi}{r^2 f  W(R^H_{\ell\omega}, R^{\infty}_{\ell\omega})},
\end{equation}
where $r_</r_>$ indicates the lesser/greater of $r$ and $r_q$ and $W(u_1,u_2)=u_1u_2'-u_2u_1'$ is the Wronskian of $u_1$ and $u_2$.  Note that the differential equation \eqref{SchwWaveEqn} guarantees that the combination $r^2 f W(u_1,u_2)$ is constant for any two homogeneous solutions $u_1$ and $u_2$, so that $N_{\ell\omega}$ is a constant normalization factor.\footnote{It is also known (e.g. \cite{schw-solns}) that the Wronskian of $R^H$ and $R^\infty$ is everywhere non-zero, so that this constant is well-defined.} This particular solution to \eqref{SchwWaveEqn} has no incoming radiation either from infinity or the horizon, so that it should correspond to the retarded Green's function in pure Schwarzschild spacetime.  This property makes it convenient for use in a computation of a self-force difference between our spacetime \eqref{bodyMetric} and the pure Schwarzschild spacetime.

We now proceed to the matching at the surface of the body, $r=r_0$.  The general (inhomogeneous) solution to equation \eqref{SchwWaveEqn} compatible with our boundary conditions may be written as the particular solution \eqref{gschw} plus a piece purely outgoing at infinity,
\begin{equation}\label{Gout}
G^{\textrm{out}}_{\ell\omega}(r) = G^{\textrm{Schw}}_{\ell\omega}(r) + C_{\ell \omega} R^\infty_{\ell\omega}(r),
\end{equation}
where $C_{\ell \omega}$ is a constant.  On the other hand, the general solution to equation \eqref{ArbWaveEqn} compatible with our boundary conditions is
\begin{equation}\label{Gin}
G^{\textrm{in}}_{\ell \omega}(r) = D_{\ell \omega} I_{\ell \omega}(\tilde{r}),
\end{equation}
where $D_{\ell \omega}$ is a constant, and $I_{\ell \omega}$ is the unique\footnote{The Frobenius method (see, e.g., \cite{boyce-diprima}) applied to equation \eqref{ArbWaveEqn} at the regular singular point $r=0$ implies that a unique (up to normalization) regular solution exists for $\Psi$ and $\Lambda$ analytic at zero.} (up to normalization)  solution to equation \eqref{ArbWaveEqn} regular at the origin.  The matching conditions \eqref{matchr0G} and \eqref{matchr0Gp} determine the constants $C_{\ell\omega}$ and $D_{\ell\omega}$.  Out interest is in $C_{\ell\omega}$, which we determine to be
\begin{equation}\label{Cchi}
C_{\ell \omega} = \left. N_{\ell\omega} R^\infty_{\ell\omega}(r_q)\frac{W(I_{\ell\omega},R^H_{\ell\omega})-\xi f^{-1}  \chi I_{\ell\omega}R^H_{\ell\omega}}{W(I_{\ell\omega},R^\infty_{\ell\omega})-\xi f^{-1} \chi I_{\ell\omega}R^\infty_{\ell\omega}} \right|_{r=r_0}.
\end{equation}
In writing this expression we have assumed that the denominator in the fraction is non-zero.  If the denominator does vanish, then our matching conditions (which follow from the equation for $G_{\ell \omega}$) become inconsistent.  Since the retarded Green's function must exist in our globally hyperbolic spacetime \eqref{bodyMetric}, this indicates that the mode-sum decomposition \eqref{greens-decomp} is not defined in that spacetime.  We show in the appendix that the denominator can vanish only for $\omega=0$, in which case it corresponds to a linear instability of the spacetime.  In the Wronskians present here and below, all derivatives are to be taken with respect to $r$ (rather than $\tilde{r}$).

Now consider the Green's function difference $\Delta G$ (defined only for $r_q>r_0$) between the general metric \eqref{bodyMetric} and pure Schwarzschild spacetime.  Equation \eqref{Gout} shows that this difference is controlled by the constant $C_{\ell\omega}$, so that we can write 
\begin{equation}\label{Gdiff}
\Delta G_{\ell\omega} = N_{\ell\omega}K_{\ell\omega}(r_0)R^{\infty}_{\ell\omega}(r)R^{\infty}_{\ell\omega}(r_q),
\end{equation}
where
\begin{equation}\label{Kl}
K_{\ell \omega} = \left. \frac{W(I_{\ell\omega},R^H_{\ell \omega})-\xi f^{-1}  \chi I_{\ell\omega}R^H_{\ell\omega}}{W(I_{\ell\omega},R^\infty_{\ell\omega})-\xi f^{-1}  \chi I_{\ell\omega}R^\infty_{\ell\omega}} \right|_{r=r_0}
\end{equation}
is a constant (depending only on $r_0$).  Equations \eqref{Gdiff} and \eqref{Kl} are the main results of this section, giving the ``Green's function difference'' $\Delta G$ mode by mode.  Integrating the Green's function difference over a given orbit via equation \eqref{sfdiff-greens} gives the difference in self-force between a body orbiting a static spherically symmetric body (equation \eqref{bodyMetric}) and the same body on the same orbit about a Schwarzschild black hole of the same mass.  As is made manifest in these results, this quantity can be computed directly from the mode functions $I_{\ell\omega}$,$R^H_{\ell\omega}$,$R^\infty_{\ell\omega}$ of the spacetimes, without the need for regularization.  As previously discussed below \eqref{Cchi}, our expression for $K_{\ell\omega}$ is well-defined in all spacetimes (of the form \eqref{bodyMetric}) for which the mode decomposition \eqref{greens-decomp} is defined.

\subsection{Expression for two non-black-hole central bodies}

Suppose one is interested is the self-force difference for two different central bodies $A$ and $B$ (rather than for a body and a black hole).  The most straightforward way to compute this difference is simply to use equations \eqref{Gdiff} and \eqref{Kl} to obtain the difference from Schwarzschild for each central body, and then subtract.  However, a more illuminating expression for the difference between $A$ and $B$ can be obtained as follows.  Define $A_{\ell\omega}$ to be the solution for central body $A$ that is regular at the origin but now extended throughout the whole spacetime \eqref{bodyMetric} by matching at $r=r_0$.  Thus $A_{\ell \omega}$ agrees with $I_{\ell\omega}$ interior to the body and becomes a particular combination of $R^H_{\ell\omega}$ and $R^\infty_{\ell\omega}$ exterior to the body.  By exactly the same computation that leads to equation \eqref{gschw}, the solution $G^A_{\ell\omega}$ regular at the origin and outgoing at infinity is written (for $r>r_0$) as
\begin{equation}
G^A_{\ell\omega}=-4\pi\frac{A_{\ell\omega}(r_<) R^{\infty}_{\ell\omega}(r_>)}{ r^2 f W(A_{\ell\omega}, R^{\infty}_{\ell\omega})},
\end{equation}
where again the denominator is known to be constant by the properties of \eqref{SchwWaveEqn}, and vanishes only in the presence of instabilities (see appendix and note at the end of this subsection).  Defining the analogous quantities $B_{\ell\omega}$ and $G^B_{\ell\omega}$ for central body $B$, we arrive at the analogous expression.  The difference $A-B$ is then given by
\begin{equation}\label{enroute}
\Delta G^{AB}_{\ell\omega} = -4\pi R^\infty_{\ell\omega}(r_>) \left( \frac{A_{\ell\omega}(r_<)}{r^2 f W(A_{\ell\omega},R^\infty_{\ell\omega})} - \frac{B_{\ell\omega}(r_<)}{r^2 f W(B_{\ell\omega},R^\infty_{\ell\omega})} \right)
\end{equation}
Since $r^2 f W(A_{\ell\omega},R^\infty_{\ell\omega})$ is known to be constant, we may choose any radius $r$ to evaluate it (and likewise for $B$).  In particular, we may choose $r=r_<$, which enables us to use the property of the Wronskian that $u_1 W(u_2,u_3) - u_2 W(u_1,u_3) = u_3 W(u_2,u_1)$ to simplify equation \eqref{enroute} to
\begin{equation}\label{nice}
\Delta G^{AB}_{\ell\omega} = -N^{AB}_{\ell\omega} R^\infty_{\ell\omega}(r) R^\infty_{\ell\omega}(r_q), \qquad N^{AB}_{\ell\omega} = 4\pi \frac{W(B_{\ell\omega},A_{\ell\omega})}{r^2 f W(A_{\ell\omega},R^\infty_{\ell\omega}) W(B_{\ell\omega},R^\infty_{\ell\omega})},
\end{equation}
where $N^{AB}_{\ell\omega}$ is a constant.  This form shows that the difference in self-force induced by central bodies $A$ and $B$ is controlled by the Wronskian of their origin-regular mode functions $A$ and $B$.  In particular, when central bodies $A$ and $B$ are identical, then the mode functions $A_{\ell\omega}$ and $B_{\ell\omega}$ will be linearly dependent, and the self-force difference vanishes by the Wronksian's vanishing on linearly-dependent functions.  Equation \eqref{nice} also has the aesthetic advantage that it makes no direct reference to the ``horizon'' Schwarzschild solution $R^H_{\ell \omega}$, which should not play a privileged role in a self-force difference between two non-black-hole spacetimes.  However, this formula offers no practical benefit over simply computing the difference from pure Schwarzschild by \eqref{Kl} for each spacetime $A$ and $B$, since the relationship of $A_{\ell\omega}$ (or $B_{\ell\omega}$) to the corresponding solutions $I_{\ell\omega}$ to the interior wave equation restores the complicated dependence on $R^H_{\ell\omega}$ (or on some other choice of a second linearly independent solution to the Schwarzschild wave equation).  Specifically, one has
\begin{equation}\label{Al}
A_{\ell\omega} = E_{\ell\omega}R^H_{\ell\omega} + F_{\ell\omega}R^\infty_{l \omega},
\end{equation}
where
\begin{align}
E_{\ell\omega} & = \left. \frac{W(I_{\ell\omega},R^\infty_{\ell\omega})-f^{-1}\xi\chi I_{\ell\omega}R^\infty_{\ell\omega}}{W(R^H_{\ell\omega},R^\infty_{\ell\omega})} \right|_{r=r_0} \label{El} \\
F_{\ell\omega} & = \left. -\frac{W(I_{\ell\omega},R^H_{\ell\omega})-f^{-1}\xi\chi I_{\ell\omega}R^H_{\ell\omega}}{W(R^H_{\ell\omega},R^\infty_{\ell\omega})} \right|_{r=r_0}. \label{Fl}
 \end{align}
Notice the simple relationship to the constant $K_{\ell\omega}$ (equation \eqref{Kl}) controlling the self-force difference between $A$ and Schwarzschild; we have  $K_{\ell\omega}=-F_{\ell\omega}/E_{\ell\omega}$.  In particular equation \eqref{Kl} is ill-defined when $E_{\ell\omega}=0$, which by \eqref{Al} indicates linear dependence of $A_{\ell\omega}$ and $R^\infty_{\ell\omega}$.

\subsection{Electromagnetic and Gravitational Cases}

We now discuss the generalization of the above treatment of self-force differences to the electromagnetic and gravitational cases.  The electromagnetic case proceeds in complete parallel with the scalar case.  The electromagnetic analogs of \eqref{sf} and \eqref{greens} are \cite{dewitt-brehme,quinn-wald,poisson}
\begin{equation}\label{sf-EM}
F^\mu = e^2 \left[ \frac{2}{3} (\dot{a}^\mu - a^2 u^\mu) + \lim_{\epsilon \rightarrow 0^+} \int_{-\infty}^{\tau-\epsilon} u_\nu \nabla^{[\nu} G^{\mu] \mu '}(z(\tau),z(\tau'))u_{\mu'}(\tau') d\tau' \right]
\end{equation}
and
\begin{equation}\label{greens-EM}
\left[ \delta^\alpha_{\ \beta} g^{\mu \nu}\nabla_\mu \nabla_\nu - R^{\alpha}_{\ \beta} \right] G^\beta_{\ \alpha '}(x,x') = -4 \pi \delta^\alpha_{\ \alpha '}\delta^{(4)}(x,x'),
\end{equation}
where the retarded Green's function should be chosen for $G^\beta_{\ \alpha '}$.\footnote{Note that a vector potential constructed from $G^\alpha_{\ \alpha'}$ using a conserved current will satisfy the Lorenz gauge condition, $\nabla_\mu A^\mu = 0$.}  These equations give the self-force on a particle of charge $e$ moving through a vacuum region of spacetime.  As in the scalar case, consideration of a self-force difference allows one to drop the limit in equation \eqref{sf-EM}, producing the analog of equation \eqref{sfdiff-greens}.  One can then follow the procedure used in section \ref{sec:arb} to derive the analogs of (\ref{greens-decomp}-\ref{Fl}).  However, the separation of the equations is considerably more complicated than the scalar case, since (e.g.) vector spherical harmonics must be used to avoid mode-mode coupling.  Since we perform no explicit calculations in the time-dependent case, we do not display these equations.  Instead, in the following section we will develop the static case separately, where scalar harmonics can be used and the equations are considerably simpler.

As in the scalar case, the electromagnetic self-force formalism is based on the wave equation on a fixed curved spacetime, so that if the spacetime is non-vacuum, the results only apply if the matter is not coupled to the electromagnetic field.  While in the scalar case it seems fair to simply declare any matter to be uncoupled at a fundamental level, in the electromagnetic case this would eliminate most known types of matter.  Instead, the restriction on the type of matter in the electromagnetic case is best stated as the requirement of a nonpolarizable medium with no net charge.  Thus, our electromagnetic results are valid only for nonpolarizable central bodies (with no net charge).  The inclusion of polarizability is not difficult at a theoretical level---one would simply replace equation \eqref{greens-EM} with a ``macroscopic'' Maxwell equation or other choice of coupled Maxwell and matter equations---but obtaining solutions for such equations may be more difficult.\footnote{An exception is the case of a conducting central body, where the matter model consists in the demand of zero electric field inside the body, which effectively contributes only a boundary condition to Maxwell's equation.  Indeed, this case was considered by Shankar and Whiting \cite{shankar-whiting} (who also consider an insulating Schwarzschild star), using a technique similar to ours.}  We note, however, that in this case the self-force formalism would automatically include backreaction effects due to polarization induced by the charge; there is no clean distinction between these and other self-force effects.

The theory of gravitational self-force \cite{mino-sasaki-tanaka, quinn-wald, poisson, gralla-wald, pound} is also highly analogous to the scalar and electromagnetic cases, with the Linearized Einstein equation playing the role of the scalar wave equation / Maxwell's equation, and the perturbed geodesic equation ``force'' playing the role of the scalar force / Lorentz force.  The main difference in the gravitational case is that the background spacetime is assumed to be vacuum.  This is done because the assumption of having matter that does not couple to the field is no longer permissible, since gravity couples to everything.  As explained and developed in \cite{pfenning-poisson}, gravitational self-forces in the presence of matter require one to consider (linearized) coupled matter and Einstein equations.  Thus, while the theory of gravitational self-force differences for static spherically symmetric bodies can be developed in analogy with the scalar and electromagnetic cases, it would require the use of Green's functions for (linearized) coupled matter and gravitational equations.

\section{Arbitrary Interior: Static case}\label{sec:arb}

The case of a charge held static at a fixed radius $r=r_q$ outside the body provides enough simplification to obtain some general results (valid for a general interior metric).  The worldline of the charge is given by $u_\alpha=(\sqrt{f(r_q)},0,0,0)$, whence equation \eqref{sfdiff-greens} gives the self-force difference to be
\begin{equation}\label{static-force}
\Delta F_r = q^2 \left. \left( \sqrt{f} \partial_r \Delta G(r,r_q,\omega=0) \right) \right|_{r=r_q},
\end{equation}
with all other components vanishing.  However, Wiseman \cite{wiseman} has shown that the self-force on a static scalar charge outside a Schwarzschild black hole vanishes.  Therefore, in the static case the self-force difference from Schwarzschild $\Delta F_r$ is in fact equal to the self-force $F_r$ (and we will drop the $\Delta$ below).  Notice that only the $\omega=0$ mode of the Green's function contributes to the static self-force.  Accordingly, $\omega=0$ will be implicit for the remainder of this section, and we will drop the label $\omega$ on the mode functions.  For $\omega=0$ the Schwarzschild mode functions are known analytically \cite{wiseman,burko-liu-soen}; they are (up to normalization) simply the Legendre functions $P_\ell(\bar{r})$ and $Q_\ell(\bar{r})$ \cite{arfken-weber} of the Schwarzschild harmonic coordinate $\bar{r}=r/M-1$.  Using the fact that $W(P_\ell(x),Q_\ell(x)) = 1/(1-x^2)$, equations \eqref{Gdiff} and \eqref{Kl} become
\begin{equation}\label{Gdiff-static}
\Delta G_\ell = -\frac{4\pi}{M} K_\ell(r_0)Q_\ell(\bar{r}_q)Q_\ell(\bar{r})
\end{equation}
and
\begin{equation}\label{K-static}
K_\ell = \left. \frac{W(I_\ell,P_\ell)-\xi f^{-1}  \chi I_\ell P_\ell}{W(I_\ell,Q_\ell) - \xi f^{-1}  \chi I_\ell Q_\ell} \right|_{r=r_0},
\end{equation}
where in equation \eqref{K-static} and below it is understood that the Legendre functions are evaluated at $\bar{r}(r)$ (but that their derivatives in the Wronskian are taken with respect to $r$).  Using equations \eqref{greens-decomp} and \eqref{static-force}, the self-force (difference) is seen to be
\begin{equation}\label{fr-static}
\Delta F_r = F_r = -\frac{q^2}{M^2}\sqrt{f(r_q)} \sum_{\ell=0}^{\infty}(2 \ell +1)
K_\ell(r_0) Q_\ell(\bar{r}_q) Q_\ell'(\bar{r}_q),
\end{equation}
where the prime indicates differentiation with respect to $\bar{r}$, and we have used the addition theorem \cite{jackson} to perform the sum over $m$.  Equations \eqref{fr-static} and \eqref{K-static} give the self-force on a scalar charge held static outside an arbitrary central body of the form \eqref{bodyMetric}.

Now consider the ``far-field'' limit in which the charge is taken far away from the body, $r_q \rightarrow \infty$.  For large $x$, the Legendre function $Q_\ell(x)$ behaves as $x^{-(\ell +1)}$, so that the $\ell =0$ mode dominates the far-field self-force.  Specifically, we have
\begin{equation}\label{frfar}
F_r = \frac{q^2M}{r_q^3} \sqrt{f(r_q)} \left. \frac{W(I_0,P_0)-\xi f^{-1}  \chi I_0 P_0}{W(I_0,Q_0) - \xi f^{-1}  \chi I_0 Q_0} \right|_{r=r_0} + \mathcal{O}\left(\frac{q^2M^3}{r_q^5} \right).
\end{equation}
We now prove that the first term in equation \eqref{frfar} vanishes when $\xi=0$.  Since $P_0$ is simply a constant (non-zero) function, its Wronskian with $I_0$ vanishes if and only if $I_0$ is a constant function.  Now, $I_0$ is the unique (up to normalization) solution to equation \eqref{ArbWaveEqn} with $\ell =\omega=0$ that is regular at the origin.  When $\xi=0$, all terms proportonal to $G_{l\omega}$ in equation \eqref{ArbWaveEqn} vanish and the equation admits a constant solution, showing that $I_0$ is constant and the Wronskian $W(I_0,P_0)$ vanishes.  Since the second term in the numerator of \eqref{frfar} also vanishes when $\xi=0$, the leading order far-field self-force vanishes when $\xi=0$.  Thus we have the following result:
\begin{itemize}
\item For minimally coupled fields, the leading order far-field self-force on a scalar charge held static outside an arbitrary static, spherically symmetric central body is  $\mathcal{O}(q^2 M^3 / r_q^5)$.  By contrast the force for non-minimally coupled fields is $\mathcal{O}(q^2 M / r_q^3)$.
\end{itemize}
The result in the minimally-coupled case may be viewed as stating that the self-force is independent of the central body (always taking the value of zero) at $\mathcal{O}(q^2 M/r_q^3)$, confirming the suggestion of figure \ref{fig:sfresults}.  Similarly, we have shown that the corresponding force in the nonminimally-coupled case is dependent in detail on the central body composition.  We note that some similar results have been obtained in \cite{anderson-fabbri} for expectation values of products of quantized scalar fields on static spherically symmetric spacetimes.

We now consider the electromagnetic analog of the above calculations.  Since the calculation is highly analogous, we simply sketch the procedure and present the results.  For a static worldline $z(\tau)$ on our spacetime \eqref{bodyMetric}, the only contribution of the Green's function to the self-force \eqref{sf-EM} is from the time-time component, which is scalar under spatial rotations.  We may therefore expand in ordinary scalar harmonics, as in the case of a scalar charge treated in section \ref{sec:arb}.  The treatment then proceeds in precise analogy, and we find that for an electric charge held fixed outside an arbitrary (nonpolarizable) body of the form \eqref{bodyMetric}, the self-force difference from pure Schwarzschild is given by
\begin{equation}\label{Fr-EM}
\Delta F_r  = -\frac{q^2}{M^2}\sqrt{f(r_q)} \sum_{\ell=0}^\infty (2\ell+1)\hat{K}_ \ell(r_0)\hat{Q}_ \ell(\bar{r}_q) \hat{Q}_\ell'(\bar{r}_q), \qquad 
\end{equation}
with
\begin{equation}\label{Kl-EM}
\hat{K}_ \ell = \left.\frac{W(\hat{I}_ \ell,\hat{P}_ \ell)}{W(\hat{I}_ \ell,\hat{Q}_ \ell)}\right|_{r=r_0},
\end{equation}
and where (as always) derivatives in the Wronskian are taken with respect to $r$ (rather than $\bar{r}$ or $\tilde{r}$).  Here $\hat{P}_\ell $ and $\hat{Q}_\ell$ are the electrostatic mode functions for Schwarzschild spacetime (i.e., the radial parts of the pure multipole solutions to the electrostatic Maxwell equation on Schwarzschild spacetime), given by derivatives of Legendre functions  \cite{cohen-wald},
\begin{align}
\hat{P}_ \ell(\bar{r}) &= \begin{cases} 1&\text{$\ell =0$} \\ \frac{1}{\ell(\ell +1)} (\bar{r}-1)P_\ell'(\bar{r}) & \text{otherwise}\end{cases} \label{Phat} \\
\hat{Q}_ \ell(\bar{r}) &= (\bar{r}-1)Q_\ell'(\bar{r}), \label{Qhat} 
\end{align}
and $\hat{I}_\ell$ is the origin-regular electrostatic mode function for the interior spacetime, given by the (unique up to normalization) origin-regular solution of
\begin{equation}\label{Ihat}
\hat{I}_ \ell''(\tilde{r}) + \left( - \Lambda'(\tilde{r}) + \frac{2}{\tilde{r}} - \Psi'(\tilde{r}) \right) \hat{I}_ \ell'(\tilde{r}) - e^{2\Lambda} \frac{\ell(\ell +1)}{\tilde{r}^2} \hat{I}_ \ell(\tilde{r}) = 0.
\end{equation}
Equations \eqref{Fr-EM}, \eqref{Kl-EM}, and \eqref{Ihat} are the electromagnetic analogs of equations \eqref{fr-static}, \eqref{K-static}, and \eqref{ArbWaveEqn} (specialized to $\omega=0$), respectively.  Note the change in sign of the $\Psi'$ term in equation \eqref{Ihat} from its scalar analog \eqref{ArbWaveEqn}.  Note also the lack of Ricci terms in equation \eqref{Ihat} (and corresponding lack of a boundary curvature term in equation \eqref{Kl-EM}), despite the presence of Ricci terms in the equation \eqref{greens-EM}.

In the far-field limit $r_q \rightarrow \infty$, again the $\ell =0$ mode dominates the self-force difference on account of the $r^{-(\ell +1)}$ behavior of $\hat{Q}_ \ell $.  However, as in the minimally-coupled scalar case, both $\hat{P}_0$ and $\hat{I}_0$ are just constants, so that the Wronskian $W(\hat{I}_0,\hat{P}_0)$ vanishes and the coefficient of the leading order term is in fact zero.  Thus the self-force difference vanishes at leading order, and we have shown that the self-force on a static electric charge is independent of the central body type at leading order in the far field, confirming the suggestion of table \ref{fig:sfresults}.

If we restrict momentarily to smooth central bodies (i.e. those with no boundary layer at $r=r_0$), then in each case investigated so far we have found that the leading-order far-field static self-force difference is given by the Wronskian of the $\omega= \ell =0$ origin-regular interior solution with the $\omega= \ell =0$ horizon-regular Schwarzschild solution.  Since the calculations done here in the scalar and electromagnetic cases generalize straightforwardly to other separable wave equations, it is clear that this conclusion must apply quite generally.  When asking if the far-field static self-force will be independent of the central body for a given wave equation, therefore, one may ask if the regular static spherically symmetric solution for any central body always agrees with that of the Schwarzschild metric.  For example, in the case of a massive scalar field, it is easy to see that the origin-regular spherically symmetric solution will be different in different spherically symmetric spacetimes, so that weak locality will not hold in this limit.  This is clearly the general case; the minimally-coupled scalar and electromagnetic wave equations have the quite special property that the regular static spherically symmetric solution is a constant for all metrics.  Therefore, it seems natural to view the weak locality of the static far-field force as a special property of the minimally-coupled scalar and electromagnetic cases.\footnote{We note that the gravitational self-force does not make sense in the static case, since no non-gravitating strut is available to hold a mass in place.  If one simply computes the field of a static charge, this manifests itself as a conical singularity in the solution \cite{keidl-friedman-wiseman}.} 

\section{Thin-shell Spacetime: Static and Circular orbits}\label{sec:shell}

We now adopt the specific choice of a central body composed only of a thin shell of matter, which enables the calculation of concrete results.  That is, we choose the interior spacetime to be flat, so that the full metric is
\begin{equation}\label{shellMetric}
ds^2 = \begin{cases} -f(r_0) dt^2 + d\tilde{r}^2 + \tilde{r}^2 d\Omega^2 & 0<\tilde{r}<r_0 \\ -fdt^2 + f^{-1}dr^2+r^2d\Omega^2 & r>r_0\end{cases}.
\end{equation}
In the language of equation \eqref{bodyMetric}, we have chosen $e^{2\Psi}=f(r_0)$ and $e^{2\Lambda}=1$.  For these choices, equation \eqref{rrtilde} becomes
\begin{equation}
\tilde{r} = r_0 + (r-r_0)f(r_0)^{-1/2}.
\end{equation}
It is easy to see that while the metric components in the coordinates $t,r,\theta,\phi$ are continuous functions across the boundary $r=r_0$, their $r$-derivatives are not.  Their second $r$-derivatives therefore contain a delta function, which is interpreted as the spacetime curvature due to a thin shell of matter present at $r=r_0$ \cite{israel,poisson}.  Our interest is in the Ricci scalar of the spacetime, which we calculate to be
\begin{equation}
R = \chi \delta(r-r_0)
\end{equation}
with
\begin{equation}\label{chichi}
\chi = \frac{2}{r_0}\left( 2 \sqrt{f(r_0)} - 2 f(r_0) - \frac{M}{r_0}\right).
\end{equation}
The quantity $\mathcal{T} \equiv - (1/8\pi) f^{-1/2} \chi$ has the interpretation of the trace of the surface stress-energy tensor of the shell, with the factor of $f^{-1/2}$ correcting for the fact that $\chi$ is the coefficient of a delta function in $r$, rather than in proper distance.  For $r_0 \rightarrow \infty$ we have $-\mathcal{T} \rightarrow (1/4\pi) M/r_0^2$, giving the surface density of a Newtonain shell.  For $r_0 \rightarrow 2M$, we have $\mathcal{T} \rightarrow \infty$, showing that there cannot be a shell as small as its Schwarzschild radius.  Finally, at $r_0=(9/4)M$ we have $\mathcal{T}=0$.  It is curious that this occurs at $r_0=(9/4)M$, which is the minimum size established by Buchdahl \cite{buchdahl} for smooth interiors.  (For example, a constant density star develops a pressure singularity for $r_0/M=9/4$.)  For our purposes the relevance of $r_0=(9/4)M$ is as 1) the size of shell where the force is independent of the coupling to curvature and 2) the maximum size for which instabilities can occur for $\xi>0$ (see appendix).  We note that our result for the Ricci scalar agrees with the (more general) results of \cite{eid}.

A major advantage of the thin-shell spacetime is that---since the interior metric is flat---its mode functions are known in closed form, even in the non-static case.  Equation \eqref{ArbWaveEqn} for the mode functions becomes simply
\begin{equation}
G_{\ell\omega}''(\tilde{r}) +\frac{2}{\tilde{r}}G_{\ell\omega}'(\tilde{r}) -  \left(\frac{\ell(\ell +1)}{\tilde{r}^2} - f(r_0)^{-1} \omega^2 \right) G_{\ell\omega}(\tilde{r})  = 0,
\end{equation}
which is of course the ordinary scalar wave equation for flat spacetime.  The solution regular at the origin is simply (e.g. \cite{jackson})
\begin{equation}\label{Ibessal}
I_{\ell\omega}(r) = \begin{cases} \tilde{r}^ \ell & \omega=0 \\ J_ \ell\left(\omega \tilde{r} f(r_0)^{-1/2} \right)& \omega \neq 0 \end{cases},
\end{equation}
where $J_\ell(x)$ gives the spherical Bessel function regular at $x=0$.

\subsection{Static Case}

\begin{figure}[t]
\includegraphics[width=140mm]{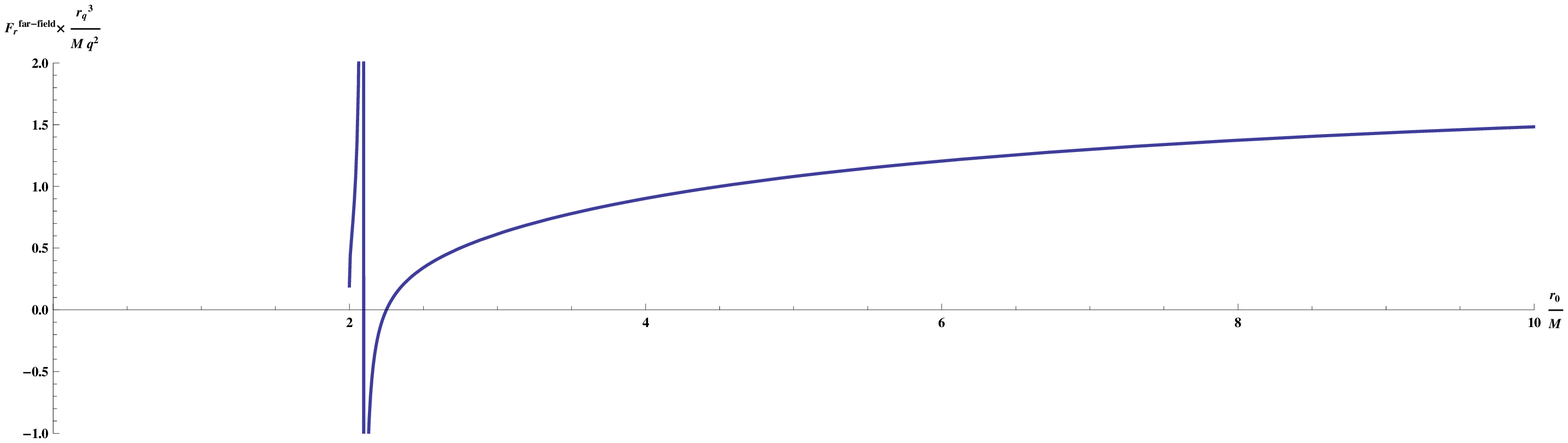}
\caption{A plot of the coefficient of the far-field force, equation \eqref{frfarshell}, for $\xi=1$ as the shell radius $r_0$ is varied.  For very large $r_0$ (not shown), the coefficient asymptotes to $2$, the Newtonian result.  For $r_0$ close to $2M$, the coefficient approaches the black hole result of zero (albeit with infinite slope).  The force changes sign at $r_0=2.25M$, and becomes arbitrarily large near an instability (see appendix) present for $r_0 \approx 2.09 M$.}
\label{fig:staticplot}
\end{figure}

With exact solutions for the interior mode functions at hand, we can evaluate \eqref{fr-static} explicitly to determine the self-force on a static charge.  We reproduce the results of \cite{burko-liu-soen} for $\xi=0$, and find that $\xi>0$ increases the repulsive self-force.  (We exclude $\xi<0$, which can give rise to instabilities (see appendix) even for large shells.  We also exclude $r_0<2.25M$ for $\xi>0$ to avoid instabilities.)  We note that the self-force diverges as the particle approaches the shell, as found for minimal coupling in \cite{burko-liu-soen}.  Our main interest in the static case, however, is in the far-field self-force \eqref{frfar}.  Using the explicit forms $I_0=1$,$P_0=1$, and $Q_0(x)=1/2 \log[(x+1)/(x-1)]$, the leading order piece of the self-force becomes
\begin{equation}\label{frfarshell}
F^{\text{far-field}}_r =  \frac{q^2M}{r_q^3} \frac{2 \xi r_0^2 \chi(r_0) }{2M + \xi r_0^2 \chi(r_0) \log \left(\frac{r_0}{r_0-2M}\right)}.
\end{equation}
Equation \eqref{frfarshell} gives the self-force on a static charge held a large distance away from a spherical shell of radius $r_0$.  Two limits are of interest.  First, when the shell radius $r_0$ becomes very large, the results should go over to those of Pfenning and Poisson \cite{pfenning-poisson} for Newtonian bodies.  Indeed, it is easy to see that we have
\begin{equation}\label{frfarshell-newt}
F^{\text{far-field}}_r \rightarrow 2 \xi \frac{q^2 M}{r_q^3} \qquad \qquad \textrm{as} \ \ r_0 \rightarrow \infty,
\end{equation}
in agreement with their results.  Second, when the shell radius $r_0$ is very nearly $2M$, the result should approach Wiseman's result \cite{wiseman} of zero self-force in the black hole case.  Indeed, it is easy to see that we have
\begin{equation}\label{frfarshell-bh}
F^{\text{far-field}}_r \rightarrow 0 \qquad \qquad \textrm{as} \ \ r_0 \rightarrow 2M.
\end{equation}
Therefore, our results do reproduce the known results in the limiting cases of weakly curved (Newtonian) and highly curved (black hole) central bodies.  To see the regime intermediate between these extremes, we can plot equation \eqref{frfarshell} directly (figure \ref{fig:staticplot}).  This illustrates the change in far-field self-force as the spacetime transitions from one that is weakly curved everywhere to a spacetime with a highly curved central object.  For $\xi>0$, the self-force is repulsive for most of parameter space, becoming attractive only in the dubious regime $r_0<2.25M$ where instabilities and arbitrarily large self-forces can occur (see appendix).  The sensitivity of the far-field self-force to the central object type dramatically demonstrates the non-locality of the tail integral.

\subsection{Circular Orbits}

The four-velocity of a circular (equatorial) orbit of radius $r_q$ and frequency $\Omega$ is given by $u^\alpha=u^t(1,0,0,\Omega)$ with $u^t = (f(r_q)-\Omega^2r_q^2)^{-1/2}$.  For geodesic orbits (which we consider), one has the radius-frequency relationship $\Omega=\sqrt{M/ r_q^3}$, allowing one to simplify this to $u^t=1/\sqrt{1-3M/r_q}$.  The worldline coordinates are $z^\mu(\tau) = (u^t \tau, r_q, \pi/2,\Omega u^t \tau)$.  Evaluating equation \eqref{sfdiff-greens} given the decomposition \eqref{greens-decomp} then gives 
\begin{align}\label{intermezzo}
\Delta F^\mu &= \frac{q^2}{u^t}  \left(g^{\mu \nu} + u^\mu u^\nu \right) \times \nonumber \\ & \nabla_\nu \left. \left(   \sum_{\ell=0}^{\infty}\sum_{m=-\ell}^\ell K_{\ell\hat{\omega}}(r_0) {N}_{\ell\hat{\omega}} R^{\infty}_{\ell\hat{\omega}}(r)R^{\infty}_{\ell\hat{\omega}}(r_q) Z^m_\ell P_{\ell }^m(\cos \theta)  P_{\ell}^m(0) e^{im(\phi-\Omega t)}\right) \right|_{x^\mu=z^\mu},
\end{align}
where we have 
\begin{equation}
Z^m_\ell =  \frac{(2 l + 1)}{4\pi} \frac{(\ell - m)!}{(\ell + m)!}
\end{equation}
and $\hat{\omega} \equiv m\Omega$.  The integral over frequencies collapsed to a discrete sum on account of the harmonic dependence on the $t$ and $\phi$ cordinates.  Component-by-component, equation \eqref{intermezzo} gives
\begin{align}
\Delta F_t & = - \Omega \Delta F_\phi \label{Ft} \\
\Delta F_r & = \frac{q^2}{u^t} \sum_{\ell=0}^{\infty}\sum_{m=-\ell}^ \ell  {N}_{\ell\hat{\omega}} R^{\infty \prime}_{\ell\hat{\omega}}(r_q) R^{\infty}_{\ell\hat{\omega}}(r_q) Z^m_\ell P_\ell^m(0)^2 K_{\ell\hat{\omega}}(r_0) \label{Fr} \\
\Delta F_\theta & = 0 \label{Ftheta} \\
\Delta F_\phi & = \frac{q^2}{u^t} \sum_{\ell=0}^{\infty}\sum_{m=-\ell}^ \ell i m {N}_{\ell\hat{\omega}} R^{\infty}_{\ell\hat{\omega}}(r_q)^2  Z^m_\ell P_\ell^m(0)^2 K_{\ell\hat{\omega}}(r_0). \label{Fphi}
\end{align}
Note that the summands in equations \eqref{Fr} and \eqref{Fphi} complex conjugate under $m\rightarrow-m$, implying that only their real parts contribute to the overall (real) sum.

Equations (\ref{Ft}-\ref{Fphi}) give the self-force difference from pure Schwarzschild for a particle in circular orbit about an arbitrary central body of the form \eqref{bodyMetric}.  For the thin-shell spacetime, the interior mode functions $I_{\ell\hat{\omega}}$ are given by \eqref{Ibessal}, so that the only quantities not known in closed form (or readily available numerically) are then the Schwarzschild mode functions $R^\infty_{\ell m}$ and $R^H_{\ell m}$.  We therefore compute these numerically, by solving equation \eqref{SchwWaveEqn} subject to the boundary conditions \eqref{RH} and \eqref{RI}.  One uses these equations to provide initial values\footnote{In fact, we use a higher-order series representation to provide the initial values, in order to allow the integration to begin closer to the particle.  Higher-order terms in the end behavior \eqref{RH} and \eqref{RI} are computed from a recursion relation that follows from equation \eqref{SchwWaveEqn}, as described (e.g.) in \cite{detweiler-messaritaki-whiting} and \cite{gralla-friedman-wiseman}.} far from the particle, and then numerically integrates equation \eqref{SchwWaveEqn} to the required $r_q$.  We used the software package Mathematica 7 \cite{mathematica} for all of our numerical computations.  We have verified the accuracy of our mode functions by using them to compute the fluxes reported in \cite{gralla-friedman-wiseman} (which have now been independently confirmed in \cite{warburton-barack}) in the special case of Schwarzschild.  We have also verified that the flux agrees with the local dissipative self-force for all radii, in the manner described in (e.g.) \cite{gralla-friedman-wiseman}.

With all the ingredients assembled it is now a simple matter to evaluate equations \eqref{Fr} and \eqref{Fphi} for various values of $\xi,r_0,r_q$.  As in the static case, we exclude $\xi<0$ as well as $\xi>0,r_0<2.25M$ to avoid instabilities.  In general the self-force difference (radial and angular) is positive, increasing the radial force found in \cite{diazrivera-detweiler-messaritaki-whiting}, and decreasing the magnitude of the negative angular force (e.g. \cite{gralla-friedman-wiseman}).   As $\xi$ is increased (at fixed $r_0,r_q$), the (radial and angular) self-force difference increases.  Likewise, as $r_0$ is increased (at fixed $r_q,\xi$), the (radial and angular) self-force difference increases.  As in the static case, the radial self-force diverges as $r_0$ approaches $r_q$ (while the angular self-force remains finite).  As $r_q$ is increased (at fixed $\xi$,$r_0$), the magnitude of the self-force difference decreases.  Since we find no qualitative differences for $\xi>0$ (and $r_0>2.25M$), for simplicity we will restrict to minimal coupling ($\xi=0$) for the remainder of this section.

\begin{figure}[t]\tiny
\begin{center}
\includegraphics[width=100mm]{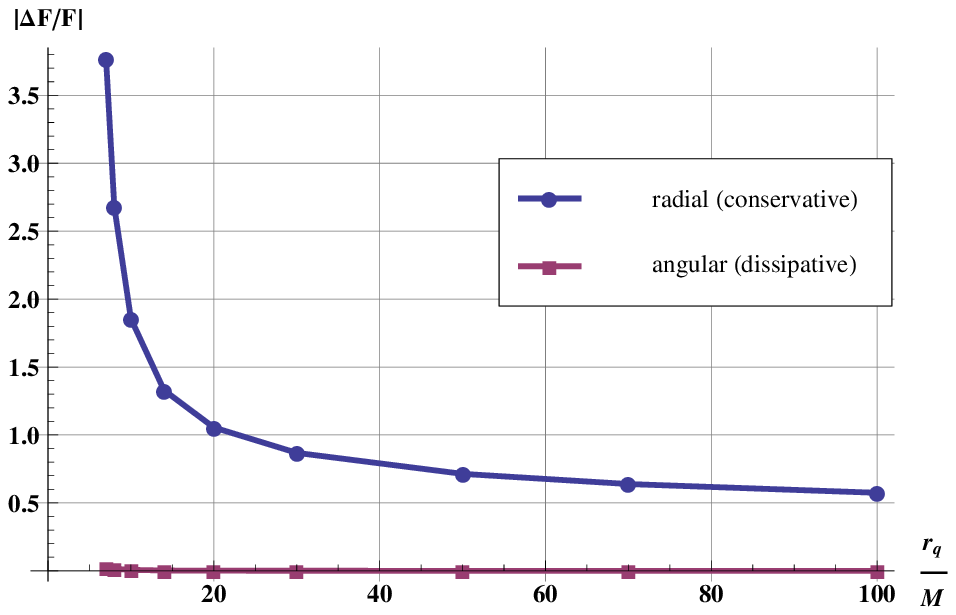}
  \begin{tabular}{ | c | c | c | c | c | c | c | c | c | c | c |}
    \hline
    \ \ \ \ \ $r_q$: & 6 & 7 & 8 & 10 & 14 & 20 & 30 & 50 & 70 & 100 \\ \hline
    radial & 7.248 & 3.772 & 2.679 & 1.856 & 1.329 & 1.055 & .8681 & .7136 & .6386 & .5742 \\ \hline
    angular & 4.229e-2 & 2.385e-2 & 1.489e-2 & 6.984e-3 & 2.233e-3 & 7.546e-4 & 2.141e-4 & 4.471e-5 & 1.611e-5 & 5.447e-6 \\ \hline
  \end{tabular}
\end{center}
\caption{The fractional self-force difference $|\Delta F/F^{\textrm{black hole}}|$, for radial (conservative) and angular (dissipative) components.  (Black hole self-force results are taken from \cite{diazrivera-detweiler-messaritaki-whiting}.) Here we have used units where $M=1$ and taken $\xi=0$, $r_0=5$.  The change in central object is seen to have a much larger effect on the radial force than on the angular force.  The increase in the radial component near the shell is associated with the divergence of the radial shell self-force when $r_q=r_0$.}
\label{fig:sffrac}
\end{figure}

Figure \ref{fig:sfnumbers} lists the results for various values of $r_0$ and $r_q$, showing some of the trends discussed above.  In order to better understand the results, however, it is instructive to instead consider the \textit{fractional} self-force difference; that is, the magnitude of the self-force difference divided by the self-force in one of the spacetimes.  The self-force on a scalar charge in circular orbit about a Schwarzschild black hole was computed in \cite{diazrivera-detweiler-messaritaki-whiting}.  The authors list only the radial self-force, but the angular self-force may be easily computed from the mode functions $R^H_{\ell\omega}$ and $R^I_{\ell\omega}$ as described (e.g.) in \cite{gralla-friedman-wiseman}.  (The radial self-force, on the other hand, requires regularization.) Using data from \cite{diazrivera-detweiler-messaritaki-whiting} for the radial black hole self-force and our own computations for the angular black hole self-force, we compute $|\Delta F_r/F_r^{\textrm{black hole}}|$ (and likewise for $\phi$) for a variety of radii (figure \ref{fig:sffrac}).  The main advantage of the fractional self-force difference is that the radial and angular components can be sensibly compared.  We see that the radial fractional self-force difference is always order unity, while the angular fractional self-force difference is small and falls off rapidly.

\begin{figure}[t]\tiny
\begin{center}
  \begin{tabular}{ | c | c | c | c | c | c | c | c | c | c | c |}
    \hline
    \ $r_0 \Big \backslash r_q $ & 6 & 7 & 8 & 10 & 14 & 20 & 30 & 50 & 70 & 100 \\ \hline
    5 ($r$) & 1.12e-3 & 2.96e-4 &1.09e-4 & 2.56e-5 & 3.61e-6 & 5.21e-7 & 6.23e-8 & 4.53e-9 & 8.20e-10 & 1.35e-10 \\ \hline
    10 ($r$) & \ & \ & \ & \ & 1.44-5 & 1.44e-6 & 1.52e-7 & 1.06e-8 & 1.90e-9 & 3.12e-10 \\ \hline
    20 ($r$) & \ & \ & \ & \ & \ & \ & 5.02e-7 & 2.50e-8 & 4.23e-9 & 6.76e-10 \\ \hline
    5 ($\phi$) & 2.24e-4 & 7.81e-5 & 3.29e-5 & 8.26e-6 & 1.13e-6 & 1.45e-7 & 1.47e-8 & 8.47e-10 & 1.31e-10 & 1.82e-11 \\ \hline
    10 ($\phi$) & \ & \ & \ & \ & 1.57e-6 & 1.92e-7 & 1.93e-8 & 1.11e-9 & 1.71e-10 & 2.38e-11 \\ \hline
    20 ($\phi$) & \ & \ & \ & \ & \ & \ & 3.02e-8 & 1.65e-9 & 2.53e-10 & 3.51e-11 \\ \hline
  \end{tabular}
\end{center}
\caption{A table of results for the self-force difference, equations \eqref{Fr} and \eqref{Fphi}, in units where $M=1$ and with $\xi=0$ and $q=1$.  The first/last three rows show the radial/angular self-force difference for three different shell radii $r_0$. The columns vary the particle radius $r_q$.}
\label{fig:sfnumbers}
\end{figure}

This result can be understood by recalling that the radial force is conservative, while the angular force is dissipative.  More specifically, with $G_{\textrm{ret/adv}}$ the retarded/advanced Green's function, one can define $G_{\textrm{diss}} = (1/2)(G_{\textrm{ret}}-G_{\textrm{adv}})$ and $G_{\textrm{cons}} = (1/2)(G_{\textrm{ret}}+G_{\textrm{adv}})$, so that we have $G_{\textrm{ret}} = G_{\textrm{diss}} + G_{\textrm{cons}}$.  A self-force computed with $G_{\textrm{cons}}$ will be time-symmetric and therefore cannot carry energy away from the system (hence the name conservative); thus the remainder is entirely responsible for energy loss (hence the name dissipative).  Noting that the interchange ret$\leftrightarrow$adv corresponds for circular orbits to $t\leftrightarrow-t,\phi\leftrightarrow-\phi$, it is easily established that the radial force is purely conservative, while the angular force is purely dissipative.  This puts the results of figure \ref{fig:sffrac} in context: It makes sense that the dissipative piece would be dominated by local emission of radiation, with only small effects due to the excitation of the central body.\footnote{Indeed, one can check that the Newtonian result of \cite{pfenning-poisson}---which explicitly takes a radiation-reaction form---gives the correct dissipative self-force (for either spacetime) at the percent-level for $r_q=30M$, at the five-percent-level for $r_q=15M$, and still within fifty percent at $r_q=6M$.}  On the other hand, there is no reason why the conservative piece should not depend strongly on the central body, especially in light of the fact that conservative self-forces often owe their entire existence to boundary conditions.  For example, the force on an electric charge held static outside a conductor would be counted as a (conservative) self-force under our definitions.

The result of figure \ref{fig:sffrac}---combined with the above interpretation---suggests that in general the dissipative part of the self-force should be more weakly local (i.e., independent of the central body type) than the conservative part, even in the strong field regime.  It would be very interesting if the dissipative self-force could also be shown to be more truly local (in the sense of tail falloff) than the conservative part.  While this cannot be true in the weak field given the computations of \cite{dewitt-dewitt,pfenning-poisson}, it seems possible that it could become true for near-field orbits.  In general, it seems plausible that the tail could become more local in the near-field, where the light reflection time is decreased.  
 
\begin{figure}[t]\tiny
\begin{center}
\includegraphics[width=100mm]{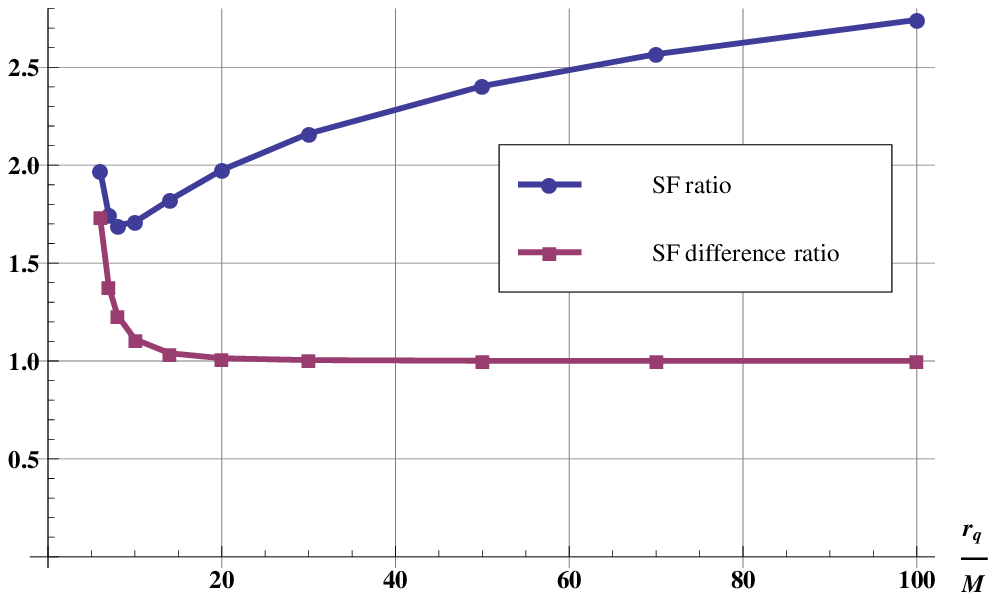}
  \begin{tabular}{ | c | c | c | c | c | c | c | c | c | c | c |}
    \hline
    \ \ \ \ \ \ \ \ \ \ \ \ \ \ \ \ $r_q$: & 6 & 7 & 8 & 10 & 14 & 20 & 30 & 50 & 70 & 100 \\ \hline
    $F_r^{\textrm{circ}}/F_r^{\textrm{static}}$ & 1.975 & 1.746 & 1.693 & 1.710 & 1.822 & 1.975 & 2.161 & 2.404 & 2.567 & 2.742 \\ \hline
    $\Delta F_r^{\textrm{circ}}/\Delta F_r^{\textrm{static}}$ & 1.735 & 1.380 & 1.232 & 1.111 & 1.039 & 1.014 & 1.004 & 1.001 & 1.000 & 1.000 \\ \hline
  \end{tabular}
\end{center}
\caption{A plot of the ratio of the circular orbit radial self-force to the static case radial self-force, along with the same quantity for the self-force difference (which agrees with the self-force in the static case, $\Delta F_r^{\textrm{static}}=F_r^{\textrm{static}}$).  We see that there is disagreement in the self-force even for large $r_q$, while there is very good agreement for the self-force difference.  The results of \cite{diazrivera-detweiler-messaritaki-whiting} have been used in the calculation of these quantities.  We have used units where $M=1$, and taken $r_0=5$, $\xi=0$.}
\label{fig:circstat}
\end{figure}

An interesting question to ask is to what extent the circular orbit conservative self-force agrees with the static case self-force.  In pure Schwarzschild spacetime there is no agreement: the static case self-force vanishes, while the circular orbit force does not.  Likewise, the static and circular orbit self-forces do not agree for the thin-shell spacetime, differing by factors of order unity.  However, the self-force \textit{difference} does in fact closely agree, as seen in figure figure \ref{fig:circstat}.  The good agreement even at reasonably small radii (at the percent level by $r_q=20M$)\footnote{If $\xi$ or $r_0$ is increased, the radius of agreement moves outward somewhat.} indicates that the circular orbit radial self-force in the shell spacetime may be estimated by taking the circular orbit results in pure Schwarzschild spacetime, and adding to them the self-force (difference) computed for the shell in the \textit{static case}.  Combined with the result that the dissipative self-force difference is very small, this suggests that (except in the very strong field) only static case results are necessary to convert black hole results to results for other bodies.

\section{Summary}
For a particle orbiting a static spherically symmetric body, we have studied the dependence of the self-force on the choice of interior metric.  We began with a general treatment of the problem in the scalar case (additionally outlining the electromagnetic and gravitational cases), and then considered specific central bodies and/or orbits.  We first considered static charges in the scalar and electromagnetic cases, confirming the suggestion of earlier work \cite{pfenning-poisson} that the far-field self-force is independent of the central body type in the minimally coupled scalar and electromagnetic cases, but dependent on the central body type in the nonminimally coupled scalar case.  We then adopted the specific choice of a thin-shell central body and computed the self-force difference from a black hole central body in the case of static and circular orbits of scalar charges.  We found that the change in central body has a much larger effect on the radial (conservative) self-force than on the angular (dissipative) self-force difference.  We also found that the radial self-force difference is well approximated by the static case (radial) self-force difference, raising the possibility of using static case results to correct self-forces (for arbitrary orbits) for a change in central body.

\begin{acknowledgments}
We thank Robert Wald for his expert guidance, as well as for several helpful comments and suggestions.  This research was supported in part by NSF grants PHY08-54807 and PHY07-55071 to the University of Chicago.
\end{acknowledgments}

\appendix
 
\section{Instabilities}

If the denominator in the expression \eqref{Kl} for the self-force difference vanishes, then that expression becomes ill-defined, indicating (as remarked below \eqref{Cchi}) that the mode decomposition \eqref{greens-decomp} is not defined.  We now show that the origin of this problem---if it occurs---is a linear instability of the offending spcetime.  Since the denominator is proportional to $E_{\ell\omega}$ of equation \eqref{El}, it is clear by equation \eqref{Al} that it vanishes only for the special case where the interior solution $I_{l\omega}$ matches directly to the infinity solution $R^\infty_{\ell\omega}$ (with zero coefficient of the horizon solution $R^H_{\ell\omega}$).  This would give a homogeneous solution to scalar wave equation on the spacetime \eqref{bodyMetric} that is bounded and purely outgoing at infinity.  If $\omega \neq 0$, it is easy to check that any such solution would violate conservation of total energy, and therefore cannot exist.  However, we can have such behavior when $\omega=0$, which corresponds to the existence of a static homogeneous solution that is bounded.  In this case multiplication by $t$ gives a second spatially-bounded homogeneous solution, so that the spacetime has a linear instability.  While it may be possible to make sense of a self-force in such a spacetime, we make no attempt in this paper.

However, it is still useful to categorize---as best as possible---when this type of instability can occur.  This is a straightforward matter for the $\ell=0$ mode of the thin-shell spacetime, for which explicit expressions were worked out in section \ref{sec:shell}, and we carry this out below.  The main conclusions are that the product $\xi \chi$ must be negative (corresponding to a negative mass-squared term in the Klein-Gordon equation) for an instability to occur; and for $\xi>0$, this can happen only for shells in the extreme region $r_0<2.25M$.  We are unable to provide analogous results for $\ell>0$, although it would seem reasonable to expect the same properties to hold.

We will use the properties of $\chi(r_0)$ (equation \eqref{chichi}) that 1) $\chi$ is bounded on $(2M,\infty)$; 2) $\chi$ has one zero at $r_0=2.25M$, being negative on $(2M,2.25M)$ and positive on $(2.25M,\infty)$; and 3) $\chi$ has one local maximum at $3M$, being increasing on $(2M,3M)$ and decreasing on $(3M,\infty)$.  Now, the denominator of equation \eqref{frfarshell} vanishes---and hence there is an $\ell=0$ instability---if and only if 
\begin{equation}\label{bingo}
-1= \xi \chi(r_0) \frac{r_0^3}{M}\log \left(\frac{r_0}{r_0-2M}\right)\equiv h(r_0).
\end{equation}
Thus, $\xi \chi$ must be negative for the instability to occur.

To analyze in more detail, consider first the case of $\xi>0$.  In this case $h(r_0)$ is negative, monotonic, and unbounded on $(2M,2.25M)$.  This implies that it intersects $-1$ exactly once in that domain.  For $r_0>2.25M$, on the other hand, $h(r_0)$ is positive and does not intersect $-1$.  Therefore, for $\xi>0$ there is exactly one radius for which an instability occurs, and this radius is bounded by $2M<r_0<2.25M$.

Now consider $\xi<0$.  This flips the sign of $h(r_0)$ so that it may intersect $-1$ only for $r>2.25M$.  Since $h(r_0)$ is bounded and contains a single local maximum (at $r_0=3M)$, this means that there may be zero, one, or two intersections with $-1$.  The value of $\xi$ for which there is one intersection is determined by $h(3M)=-1$, giving $\xi \approx -.98$.  For $-.98<\xi<0$ there is no intersection, and for $\xi<-.98$ there are two intersections.

\end{document}